\documentclass[]{revtex4}
\usepackage{amsmath}
\usepackage{amssymb}
\usepackage{graphicx}
\usepackage{bm}

\def\eqnn#1{Eq.~(\ref{eq:#1})}
\def\sect#1{Sec.~\ref{sec:#1}}

\def\figno#1{Fig.~\ref{fig:#1}}

\def\vev#1{\left\langle #1\right\rangle}

\def\cP{{\cal P}}

\def\figW{86mm}
\begin{document}
\title{
  Observation of Reflectance Fluctuations in Metals
}
\author{Takahisa   Mitsui and 
Kenichiro Aoki}
\affiliation{Research and Education Center for Natural Sciences and
  Dept. of Physics, Hiyoshi, Keio University, Yokohama 223--8521,
  Japan}
\begin{abstract}
    Through the study of the power spectra of a monochromatic light
    beam reflected by metallic mirrors, fluctuations in their
    reflectance is observed.  The power spectra were obtained down to
    a factor $10^{-6}$ below the Standard Quantum Limit, with a
    dynamic range of $10^5$ in the frequency and power, using methods
    we developed. The properties of the spectra are investigated and
    their dependence on the material is analyzed.  The physics underlying the
    phenomenon is also discussed. These fluctuations provide a new
    window into the degrees of freedom responsible for the reflection
    process in metals.
\end{abstract}
\maketitle
\section{Introduction}
\label{sec:intro}
Reflections from surfaces, such as mirrors, are ubiquitous, and are an
integral part of everyday life. In physics, studying optical
properties of materials is perhaps the most powerful tool for
investigating their electronic and  vibrational
properties\cite{JC,HOC,Ziman,AM}. As such, in metals, the subject has
been studied for some time, and continue to be studied actively to
this day\cite{Au2012,Ag2015,Tanner}.  Properties of reflection are
known to depend on the wavelength of light, temperature and the
material\cite{JC,HOC,Ziman,AM,Ujihara}, and can further depend on
geometric aspects of the material, such as its size and thickness.
However, fluctuations inherent in mirrors, on which we report here,
seem not to have been studied so far.  The problem we address may be
phrased from a different intriguing perspective --- can an ideal
mirror yield a ``perfect'' reflection?  Reducing this question to its
simplest concrete form, if we shine a monochromatic light on an ideal
metallic mirror, can we tell whether the light has been reflected or
not, just from the properties of the reflected light itself?  If so, can we tell
by what material?  The answers we find are positive for both
questions.  The underlying reason is that the reflection is caused by
microscopic degrees of freedom, such as electrons and ion
cores\cite{Ziman,AM}. All these degrees of freedom fluctuate both
thermally and quantum mechanically, so that they affect the light, at
some level. This effect should in principle be detectable, though the
question remains whether this is possible within practical limits. While
the fluctuations are indeed small, we have measured the fluctuations
in the reflectance in metallic mirrors and found their properties to
depend on the material. This opens another window into the degrees of
freedom responsible for reflection in metals.

The paper is organized as follows: In \sect{exp}, we explain the
design and the realization of the experiment to measure the
fluctuations in the reflectance of metals. The results obtained in the
measurements are explained, and analyzed in \sect{results}. The
meaning of the results and their underlying physics are discussed in
\sect{disc}.
\section{The Concept of the Experiment and Setup}
\label{sec:exp}
\begin{figure}[htbp]\centering
    \includegraphics[width=80mm,clip=true]{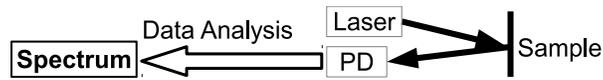}
    \caption{Basic concept of the experiment: Laser light is
      shone on the sample mirror and the reflected light is detected by a
      photodetector (PD). The output current of the photodetector is
      analyzed and its power spectrum is computed.}
\label{fig:setupSimple}
\end{figure}
When a monochromatic light beam with a constant power is shone on a
flat metallic mirror, can  the effects of the reflection be found in the
reflected light itself? 
Away from the direction determined by the law of reflection, inelastic
scattering effects, such as Brillouin and Raman effects, can be
observed, and have proven to yield important information regarding the
elastic properties of matter, as well as the electronic and
vibrational properties of atoms and molecules that constitute the
material\cite{Sandercock,Dil,Ziman,AM}.
On the other hand, the reflected light is dominated by the elastically scattered
light and its color is unchanged, so that we essentially have only the
reflected power as its property. However, its power can depend on
time, and should the microscopic degrees of freedom contributing to
the reflection fluctuate, their effects should show up in this time
dependence. 

To measure these fluctuations, conceptually, a simple experiment can
be set up as in \figno{setupSimple}. Light is shone on a mirror, and
its reflection by the mirror is detected by a photodetector
(PD). These fluctuations should be observable in the power spectrum of
the reflected light power,
\begin{equation}
    \label{eq:spectrum}
S(f) = \int_{-\infty}^\infty\!\! d\tau \,e^{-i2\pi f \tau} \vev{ \cP(t)
  \cP(t+\tau)} 
={1\over {\cal T}}\vev{\left|\tilde \cP(2\pi f)\right|^2}
\quad,    
\end{equation}
where $\cP$ is the power of the reflected light, measured by the
photodetector, and $\vev{\cdots}$ indicates the ensemble
average\cite{SpectrumRef}. $\cal T$ is the measurement time and tilde
denotes the Fourier transform.  Fluctuations in the reflectance is $
S_R(f) = S(f)/\bar\cP^2$, where $ \bar\cP$ is the average power of the
reflected light.
In reality, measurements from such an implementation are dominated by
the shot-noise\cite{SN,SNRice,SN2}, the random power fluctuations in
light due to its discrete quantum nature, often referred to as the
``Standard Quantum Limit''.  The shot-noise level appears as $2eI$, in
the photocurrent power spectrum, where $I$ is the photocurrent and $e$
is the electron charge magnitude.
It is
impossible, even in principle, to separate the signal from this noise,
with this kind of a simple setup.  The shot-noise appears in the same
manner both for the source and the reflected light, so that no effects
of the reflection process is observed in the light itself, with this
method.

\begin{figure}[htbp] \centering
        \includegraphics[width=\figW,clip=true]{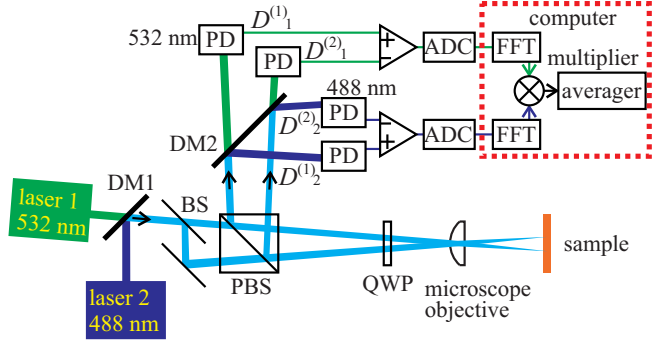}
        \caption{Experimental configuration: Differential measurements
          and averaged correlations are combined to reduce the
          unwanted noise, such as the shot-noise and the laser noise,
          in the measured spectra. DM1,2 transmit light 1 and reflect
          light 2.  Paths for the laser light 1,2 are in green, blue,
          respectively and paths common to laser light 1,2 are in
          cyan. While the light beams focused on the sample in the
          figure are well separated for illustrative purposes, the
          beams in the experiment overlap and reflect almost back
          along their original paths, but slightly shifted, due to the
          large numerical aperture (NA$=0.9$) of the objective lens.
          Fourier transforms (FFT) and averagings of the data are
          performed on a computer to obtain the spectrum (inside the
          dashed box, red).}
    \label{fig:setup}
\end{figure}
To uncover the effects of reflection, several obstacles need to be
overcome: First, unwanted noise, including shot-noise, needs to be
reduced to levels so that the fluctuations caused by the reflection
become visible. Second, it needs to be established that the observed
signal is not caused by the light causing changes to the mirror
itself, such as damaging its surface. Third, the cause of the observed
phenomenon needs to be distinguished from other possible sources of
fluctuation, such as surface waves of the
material\cite{Sandercock,Dil}.

The basic principle underlying the extraction of the spectra is to
combine the differential measurements with the averaging of the
correlated measurements. The former removes the light source noise,
which is the same, since the source is the same. The latter reduces
any noise that arises independently in the photodetector measurements,
such as the shot-noise, {\it statistically}. More concretely, two
light sources (laser 1,2), are used and each light is split into two
and shone on two locations of the sample, as seen in \figno{setup}.
These two locations are the same for the both light sources.  Accordingly,
four photocurrent measurements $D^{(\alpha)}{}_j \ (\alpha=1,2,j=1,2)$
are made, corresponding to the two focus locations on the sample, and
the two light sources. Here, $\alpha$ and $j$ label the location and
the light source, respectively. Photodetector measurements have the
following form;
\begin{equation}
    \label{eq:Daj}
    D^{(\alpha)}{}_j=S^{(\alpha)}+L_j+N^{(\alpha)}{}_j
\end{equation}
$S^{(\alpha)}$ denotes the signal, or the fluctuations, at location
$\alpha$, $L_j$ denotes the noise in the light source $j$ and
$N^{(\alpha)}{}_j$ is the shot-noise in the photocurrent
$D^{(\alpha)}{}_j$.  To obtain the spectrum, multiple measurements of
the set $\{D^{(\alpha)}{}_j\}$ are taken, and the following averaged
correlation is computed:
\begin{equation}
    \label{eq:corr}
    \vev{\overline{\left(\tilde D^{(1)}{}_1-\tilde D^{(2)}{}_1\right)}{\left(\tilde
            D^{(1)}{}_2-\tilde D^{(2)}{}_2\right)}}=
      \vev{\overline{\left(\tilde S^{(1)}-\tilde S^{(2)}+\tilde N^{(1)}{}_1-\tilde            N^{(2)}{}_1\right)}
            {\left(\tilde S^{(1)}-\tilde S^{(2)}+\tilde N^{(1)}{}_2-\tilde   N^{(2)}{}_2\right)}}
\end{equation}
Here, $\vev{\cdots}$ denotes an averaged result.  Since the
fluctuations at the different locations and the shot-noise in the
photocurrents are all independent from each other, their correlations
all go to zero statistically, in the limit of infinite number of
averagings.  Therefore, the above averaged correlation reduces
essentially to the desired spectrum,
\begin{equation}
    \label{eq:Dinf}
    \vev{\overline{\left(\tilde D^{(1)}{}_1-\tilde D^{(2)}{}_1\right)}{\left(\tilde
            D^{(1)}{}_2-\tilde
            D^{(2)}{}_2\right)}}\longrightarrow
      \vev{|\tilde S^{(1)}|^{2}}+      \vev{|\tilde S^{(2)}|^{2}}=2\vev{|\tilde S|^2}    
\end{equation}
Here, we used the property that the {\it averaged} fluctuation spectra
at the two locations on the material under identical conditions, are
the same, so that $\langle{|\tilde
  S^{(1)}|^{2}}\rangle=\langle{|\tilde
  S^{(2)}|^{2}}\rangle=\langle{|\tilde S|^{2}}\rangle$. Since
$D^{(\alpha)}{}_j$'s are photocurrent measurements, and the
photocurrent is proportional to the power of light received by the
photodetector, the spectrum \eqnn{Dinf} is essentially the power
spectrum, \eqnn{spectrum}, up to a constant.  In this averaged
correlation, the relative statistical error in the spectrum due to the
unwanted noise is the inverse of the square root of the number of
averagings.  The 
averaging of the correlation here removes any noise that is not
correlated in the two differential measurements, along with the
shot-noise. It should be noted that the averaging by itself does not
remove the shot-noise; if the differential measurement is averaged, we
obtain, in the limit of infinite number of averagings,
\begin{equation}
    \label{eq:diff}
    \vev{ |(\tilde D^{(1)}{}_1-\tilde D^{(2)}{}_1)|^2}\longrightarrow
      \vev{|\tilde S^{(1)}|^{2}}+      \vev{|\tilde S^{(2)}|^{2}}+
      \vev{|\tilde N^{(1)}|^{2}}+      \vev{|\tilde N^{(2)}|^{2}}
      =2\left(\vev{|\tilde S|^2}  +\vev{|\tilde N|^2}    \right)\quad.
\end{equation}
Similarly to \eqnn{Dinf}, the {\it averaged } noise spectrum is
independent of the location, and $\langle{|\tilde
  N^{(1)}|^{2}}\rangle=\langle{|\tilde
  N^{(2)}|^{2}}\rangle=\langle{|\tilde N|^{2}}\rangle$.  This result
contains the shot-noise that dominates the measurement, when the
signal is small, which applies to the current experimental
conditions. Therefore, the shot-noise level is determined more
precisely with more averagings, in the measurement, \eqnn{diff}.
A conceptually simpler way to reduce the relative contribution of the
shot-noise, is to increase the average light power, $\bar\cP$, since
the power spectrum, \eqnn{spectrum}, behaves as $\sim\cP^2$ and the
shot-noise behaves as $\cP$. This, in practice, is not an effective
method here --- when light powers large enough to reduce the
shot-noise to levels that uncover the spectrum are used, the sample
itself is damaged.  Furthermore, it precludes us from using smaller
light powers to systematically study the power dependence of the
spectrum, as is done in the next section.
While the observed phenomena and the measurement systems were
different, the above same basic principle, in essence, was used
previously to achieve factors of $10^{-3}$ to $10^{-5}$ reduction in
the shot-noise, in the measurements of surface thermal fluctuations of
fluids\cite{MA1,MA2}, and spontaneous noise in atomic
vapor\cite{MA3,MA4}.
%

Let us briefly mention the technical aspects of the setup 
 used in this work (\figno{setup}). Two laser sources with
wavelengths 488nm (Sapphire 488, Coherent, USA) and 532nm (Samba,
Cobolt, Sweden) were combined into a single beam with a dichroic
mirror (DM1), then split into two beams by a beam splitter (BS).  The
beams were reflected at two locations of the mirror at nearly normal
incidence (separation 77\,$\mu$m).
The light beams were focused at the mirrors down to the diffraction
limit, using a microscope objective lens (Olympus MPLFLN100XBDP) with
a high numerical aperture value (NA$=0.9$). The reason for this is
explained in the next section.
The light coming into the polarizing beam splitter (PBS) from the
source is horizontally polarized, which is then circularly polarized
at the sample using a quarter-wave plate (QWP). The reflected light is
vertically polarized by going through QWP, so that it is reflected by
PBS towards the photodetectors (PD, Hamamatsu Photonics S5973-02).
%
The reflected light powers of the beams were measured by
photodetectors, whose  differential
measurements  were digitized using  analog-to-digital
converters (ADC, Picoscope ps6404A). The digitized output was
processed on a computer to obtain the spectrum.
\section{Experimental Results and Analysis}
\label{sec:results}
\begin{figure}[htbp]\centering
    \includegraphics[width=\figW,clip=true]{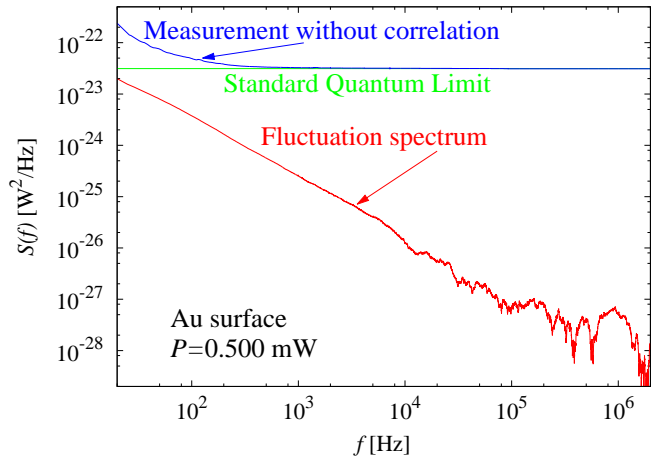}
    \caption{A typical measured fluctuation spectrum (gold surface,
      $\cP=0.5\,$mW).  Without using averaged correlations, ({\it
        c.f.\ } setup in \figno{setupSimple}), the spectrum (blue) is
      dominated by the shot-noise, or the standard quantum limit
      (green), Making use of averaged correlations, the
      fluctuation spectrum (red) was obtained down to levels $10^{-6}$ times the
      shot-noise level.  }
\label{fig:sn1}
\end{figure}
Results from a measurement using the methods in \sect{exp} are shown
in \figno{sn1}, in which it is seen that the signal is measured down
to $10^{-6}$ times the Standard Quantum Limit, with around $10^5$
factor in the dynamic range, both in the frequency and the spectral
magnitude. In the figure, measurement without correlation is the
averaged differential measurement, \eqnn{diff}, and the fluctuation
spectrum was obtained from the averaged correlation, \eqnn{Dinf}.
Here and below, the spectra were normalized using the shot-noise
level, $2eI$, in the photocurrent power spectrum.  The spectra $S(f)$
were normalized for the output signal of a single photodetector, and
the reflectance fluctuation spectra $S_R(f)$ are independent of the
normalization.
The light beam powers
applied were $8\,\mu$W to $2.5\,$mW at the mirror, per beam.  Metal coated planar
mirrors of unprotected gold (PF10-03-M03, Thorlabs, USA), unprotected
aluminum (TFAN-15S03-10, Sigma Koki, Japan), and protected silver
(PF10-03-P01, Thorlabs, USA) were used in the experiment.

The light beams in the experiment travel through, and are reflected by
various materials, including beam splitters, a quarter wave plate,
dichroic mirrors, a lens and air, apart from the sample
mirror. Therefore, it is imperative to establish that the measured
fluctuations arise from the reflections by the sample mirror at the
two beam spots.  The physics underlying this is that the beams are
focused down to the diffraction limit only at the mirrors, so that the
fluctuations from other components are averaged out over the
beam. This is why an objective lens with a large numerical aperture
was used to focus the beam to its diffraction limit at the
mirror, and another reason why a setup as simple as
\figno{setupSimple} is insufficient.
The cause of the fluctuations was also experimentally confirmed as
follows: During the measurements, the light beams from the two light
sources were focused on the same beam spots. When the beams were
focused on different points, while keeping the rest of the light paths
still overlapped, the fluctuation spectrum disappeared, showing that
the fluctuations originate from the reflections at the sample
mirror. One should add that the dependence of the spectra on the
sample material can only be explained by the the sample being the
signal source, since the experiments are otherwise identical.

Several measurements were made at different locations of the mirror to
confirm the reproducibility of the data in each case. The measurement
times for the spectra were $3\times10^4$\,s to $3\times10^5$\,s, with
more averagings, and hence longer times required for lower light
powers.

\begin{figure}[htbp]\centering
    \includegraphics[width=\figW,clip=true]{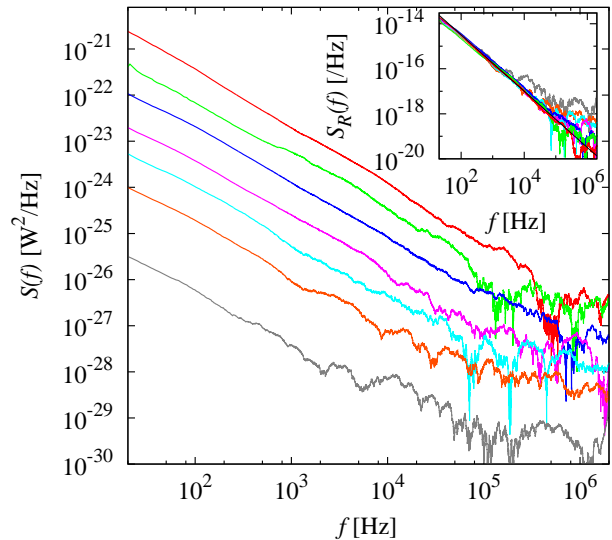}
    \caption{Measured reflection fluctuation spectra $S(f)$ for the gold
      surface for light powers, $\cP=5.00\,$mW (red), $2.54\,$mW (green),
      $1.08\,$mW (blue), $500\,\mu$W (magenta), $274\,\mu$W (cyan),
      $124\,\mu$W orange), and 
      $19.4\,\mu$W (gray). 
      Spectral magnitude increases with  $\cP$.
      (Inset) The same spectra divided by $\cP^2$ are
      seen to be identical within  experimental uncertainties, validating its interpretation
      as the reflectance fluctuation spectrum.  
     A fit $10^{-12}\times f^{-1.25}$ (black) is shown and is seen to agree
     with the reflectance fluctuation spectra nicely. 
    }
\label{fig:auSpec1}
\end{figure}
The observed signal does not exist in the incoming source light and is
the sign of reflection by the mirror. However, more work is needed to
ascertain whether this is a property of the material or the light
affecting the mirror. To this end, power spectra for the reflected
light were measured for different incoming light beam powers. The
results are shown in \figno{auSpec1} for a gold mirror. One clearly
sees that the spectra are similar in shape, which indicates that the
light is acting as a probe and is not affecting the mirror in an
essential manner. Had the light affected the mirror, it is difficult
to imagine that the spectral shape is unaffected, since the effects
should grow with the power of the light beam. Also, if the
intrinsic behavior of the mirror is being observed, the process should
be linear, so that the power spectrum should be proportional to the
square of the average light power, $\cP^2$. This can be seen in
\figno{auSpec1}(inset); by rescaling the spectra by $\cP^{-2}$,
the spectra essentially become identical, showing the similarity of
their shape and its dependence on the light power as $\cP^2$.  The
frequency dependence of the spectrum is well described over the whole
measured range by $f^{-1.25}$ ($f$ frequency), which can also be observed from the
spectra in \figno{auSpec1}.
To distinguish these fluctuations from traveling waves on the mirror
surface, previously measured in light scattering at non-specular
directions\cite{Sandercock,Dil}, the spectra were measured using
differences in the light power fluctuations at two close locations,
separated by 77\,$\mu$m.  Surface waves with longer wavelengths will
be correlated and eliminated in this differential measurement.

\begin{figure}[htbp]
    \centering
    \includegraphics[width=\figW,clip=true]{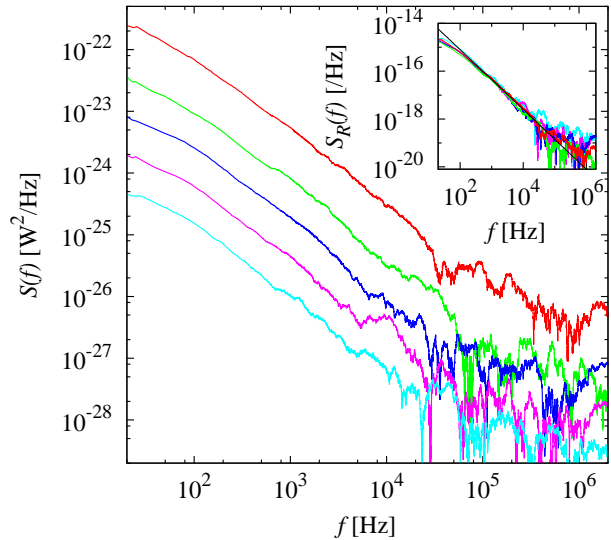}
    \caption{Measured reflection fluctuation spectra $S(f)$ for
      the aluminum surface for light powers, $\cP=2.54\,$mW (red),
      $1.08\,$mW (green), $500\,\mu$W (blue), $274\,\mu$W (magenta)
      and $124\,\mu$W (cyan). 
      Spectral magnitude increases with $\cP$.
      (Inset) The same spectra divided by
      $\cP^2$ are seen to agree, similarly to those for the gold
      mirror.  A fit $2.5\times10^{-13}f^{-1.25}$ (black) is shown and
      is seen to agree with the reflectance fluctuation spectra for
      higher frequencies, but a slight crossover behavior is seen at
      $f\sim200\,$Hz.  }
    \label{fig:alSpec1}
\end{figure}
A natural question is what happens for other metals. In
\figno{alSpec1}, the power spectra of the reflected light are shown
for an aluminum mirror for various light beam powers and in
\figno{alSpec1}(inset), the spectra rescaled by $\cP^{-2}$ are
shown. It is again seen that the spectral shapes are essentially
independent of the light power and their magnitudes behave as $\cP^2$.
The frequency dependence is again seen to be well described by
$f^{-1.25}$, but there is a slight crossover behavior at around
$200\,$Hz.   The frequency dependence of the spectra could have been
different for different metals, but interestingly enough, were similar for
gold and aluminum, except for the crossover behavior that exists for
aluminum at low frequencies.  We also measured the power spectra for
a silver mirror and found that their shapes are independent of the
power and consistent with the $f^{-1.25}$ behavior.

\begin{figure}[htbp]\centering
    \includegraphics[width=\figW,clip=true]{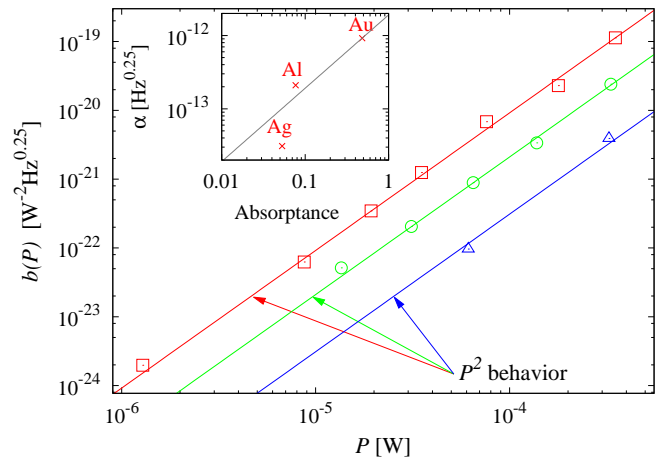}  
    \caption{The dependence of the magnitude of the spectra, $b(\cP)$,
      on the average light beam power, $\cP$, for gold ($\Box$, red),
      aluminum ($\circ$, green) and silver ($\triangle$, blue). Fits
      to the data for $\cP^2$ behavior are also shown, and are seen to
      fit the data well.  (Inset): Dependence of the reflection
      fluctuation spectra of the mirror on the material, for gold,
      aluminum, and silver (see text). Behavior proportional to the
      absorptance is also shown (gray), for reference.}
\label{fig:pDep}
\end{figure}
To quantitatively analyze the spectra, we express the spectra as
$S(f)=b(\cP)f^{-1.25}$, for $f\gtrsim200\,$Hz. The behavior of $b(\cP)$
with respect to the light power is shown in \figno{pDep}.  It is found
that $b(\cP)$ does depend on the light power as
$b(\cP)=\alpha\cP^2$, as mentioned above, 
and $\alpha$ depends on the metal. 
To understand the underlying dynamics, the dependence on the $\alpha$
on the absorptance, $A=1-R$ ($R$: reflectance), is shown in
\figno{pDep}(inset). $\alpha$ is seen to increase with 
the absorption rate, though more data is needed to establish its form
of the dependence.
\section{Underlying Physics and Discussion}
\label{sec:disc}
For optical wavelengths, the absorptance decreases with increasing
free electron density\cite{JC,HOC,AM,Ziman}, and since the statistical
noise from independent objects decrease with their number, perhaps
this suggests the source of the spectra. Free electrons play a most
important role in reflection, and it might seem that they are responsible
for the observed spectra. This is, however, unlikely for the following
reasons: First, the time scales corresponding to the observed spectra
are in the range $0.1\,$s to $0.1\,\mu$s. Free electrons in these
metals have mean free times in the order of $10^{-14}\,$s and transit
times in the light beam of $10^{-12}$\,s order.  The time scales for
free electrons are therefore too short to generate non-trivial
correlations to appear in the spectra at the observed frequency
range. Second, the energy of the photons in the light beam are
$2.4\,$eV which is of the same order as the work function for these
metals of $4\sim5\,$eV. Therefore, observation of the fluctuations in
the free electrons is expected to be non-passive and lead to nonlinear
behavior. This is inconsistent with the linear behavior seen in the
spectra, as in \figno{pDep}. Since the time scales are so short for
free electron processes, these effects should not show up in the
observed frequency range, though collective motions with the
appropriate time scales might be able to explain the observed phenomenon.

Some of the possible sources of the observed fluctuations are the
properties of the ion cores in the metal, in the light beam, down to
the skin depth. The thermal motion of the cores, or photon
interactions with the bound electrons can contribute to the observed
fluctuations.  The number density of ion cores is identical to that of
the free electrons, up to the valency factor.  The time scales of
their thermal motion are much longer than the time scales of free
electron motion, and the ion cores do not move out of the beam so that
correlations arise within the observed time scales.  It should be
noted that the acoustic vibrations\cite{Ziman,AM} have long
wavelengths compared to the separation of the beam locations, so that
they do not contribute to the spectra.
While the reflection is mainly caused by free electrons, their
fluctuations are not observable in the correlations in the range of
times that were observed here, so that the ion core contributions can
dominate the observed fluctuation spectra.
The  spectra measured in this work have behaviors close to $1/f$ over a
wide frequency range, and belong to a class often referred to as the
``$1/f$ noise'', which is widely seen in
nature\cite{Press78,DuttaHorn}. In metals, thermal motion of atoms,
including the effects of internal friction is known to lead to an
$1/f$ spectrum, with additional frequency dependencies coming from the
frequency dependency of the loss
angle\cite{Zener,Saulson90,Gillespie,Bondu}. These motions modulate
the frequency of the light through Doppler shifts, which appear in the
spectra, similarly to selective reflection\cite{Schuurmans,MS}.  The loss
angle values are not known with precision\cite{InternalFriction} and the
mechanism is technically involved.
Considering photon interactions with bound electrons, whether they can
produce photon correlations, as in \eqnn{spectrum}, the linear
response seen in \figno{pDep}, and the shape of
their spectra needs to be investigated. 

When considering thermal fluctuations of any kind to be the source of
the observed spectra, the temperature change due to light absorption
needs to be considered. If the temperature increases significantly,
the behavior becomes non-linear with respect to the light power, so
that it is not compatible with the current observations. One can
estimate the temperature rise of the sample beam spot roughly as
follows; consider a uniform semi-sphere of radius $R$, with a
semi-sphere with radius $w$ removed, at the center.  When we dissipate
heat from the inner sphere, the temperature difference $\Delta T$
between the inner and the outer boundary is $Q/(2\pi \kappa)(1/
w-1/R)$. Here, $Q$ is the heat dissipated from the inner surface and
$\kappa$ is the thermal conductivity of the material. Considering our
measurements, the absorptance of gold is almost an order of magnitude
larger than that of aluminum or silver, while their thermal
conductivities are similar, so that the gold surface may give rise to
the largest temperature difference. For the maximal beam power in our
experiments, the corresponding temperature rise is $2.7$\,K in the
above simple formula. We have also computed the temperature rise
numerically, with the experimental geometry, to find $\Delta
T=7.5\,$K. This temperature rise, while larger than the result from
the simple formula, is still much smaller than the room temperature
($\sim300\,$K), so that thermal fluctuations are quite compatible with
being the source of the observed fluctuations.

More work needs to be done to clarify the dynamics behind the
reflectance fluctuations we have observed, both qualitatively and
quantitatively. Measurements performed at different wavelengths can
lead to more information, especially since the reflection mechanism
depends on the wavelength of light\cite{HOC,Ziman,AM}. Fluctuations in
the reflectance of metals yield another window into the mechanism
underlying reflection, and understanding them would lead to deeper
insight into the degrees of freedom contributing the reflection in
metals.  Similar measurements can be performed for different metals,
other types of mirrors and various surfaces. How the spectrum depends
on the material would be an intriguing question, and even more so,
why.

\noindent{\bf Acknowledgments: }
K.A. would like to thank Keiju Murata for discussions on numerical
methods. 
K.A.  acknowledges the financial support of the Grant--in--Aid for
Scientific Research (Grant No.~15K05217) from the Japan Society for
the Promotion of Science (JSPS), and a grant  from the Keio University.

\vskip0mm\noindent
%
%

\begin{thebibliography}{99}
  \bibitem{JC} P. B. Johnson and R. W. Christy,
    Optical Constants of the Noble Metals,
    Phys. Rev. B 6, 4370--4379 (1972).
  \bibitem{HOC}E.D. Palik, Handbook of Optical Constants of Solids,
  Academic Press (New York, San Diego, 1997) 
\bibitem{Ziman}J.M.~Ziman, Theory of Solids, Cambridge University Press
  (Cambridge, UK, 1972). 
\bibitem{AM} N.W. Ashcroft and N.D. Mermin, Solid State Physics,  Saunders
  College (Philadelphia, USA, 1976).
    \bibitem{Au2012} R.L. Olmon, B. Slovick, T.W. Johnson, D. Shelton,
S-H. Oh, G.D. Boreman, and M.B. Raschke,
Phys. Rev. B 86, 235147 (2012).
\bibitem{Ag2015} H.U. Yang, J. D'Archangel, M.L. Sundheimer,
  E. Tucker, G.D. Boreman, and M.B. Raschke,
   Phys. Rev. B 91, 235137 (2015).
\bibitem{Tanner}D. B. Tanner, 
  Phys. Rev. B 91, 035123 (2015).
\bibitem{Ujihara} K. Ujihara, 
  J. Appl. Phys. 43,  2376 (1972).
\bibitem{Sandercock} J.R. Sandercock, Solid State Commun.  26,
  547 (1978).
\bibitem{Dil}
  J.G.Dil, Rep. Prog. Phys. 45, 285 (1982).
   \bibitem{SpectrumRef}See, for instance, C.W. Gardiner, {\sl
       Handbook of Stochastic Methods} (Springer-Verlag, Berlin, 1985).
\bibitem{SN}W. Schottky,  
  Ann. Phys. 57,  541 (1918).
\bibitem{SNRice} S. O. Rice, 
  Bell Syst. Tech. J.,  23, 282 (1944); 24,  46 (1945).
\bibitem{SN2} R. Loudon, The Quantum Theory of Light, 
Oxford University Press (Oxford, UK, 2000).
\bibitem{MA1}T. Mitsui, K. Aoki, 
  Phys Rev E80,   020602(R) (2009).
\bibitem{MA2}T. Mitsui and K. Aoki,
Phys. Rev. { E 87}, 042403 (2013).
\bibitem{MA3}T. Mitsui, K. Aoki,
{\sl Eur. Phys. J.} {\bf D67}, 213 (2013).
\bibitem{MA4} K. Aoki, and T. Mitsui, 
   {\sl Phys. Rev. }{\bf A 94}, 012703 (2016).
\bibitem{Press78}W. H. Press, 
  Comments Astrophys. Space Phys. 7, 103 (1978),
\bibitem{DuttaHorn}P. Dutta and P. M. Horn,
  Rev. Mod. Phys. 53, 497 (1981).
\bibitem{Zener}C.M.~Zener, Elasticity and Anelasticity of Metals, The
  University of Chicago Press (Chicago, 1948).
  \bibitem {Saulson90} P.R. Saulson, 
    Phys. Rev. D42, 2437 (1990).
\bibitem{Gillespie} A. Gillespie, F. Raab, 
  Phys. Rev.  D52, 577 (1995).
  \bibitem{Bondu}F. Bondu, J.Y. Vinet,
    Phys. Lett. A198, 74 (1995).
 \bibitem{Schuurmans}M. F. H. Schuurmans, 
   J. Phys, 37, 469 (1976).
 \bibitem{MS}
   T.Mitsui,  K.Sakurai, Jpn. J. Appl. Phys. 36, 896 (1997).
\bibitem{InternalFriction}  M.S. Blanter, I.S. Golovin,
  H.~Neuh\"auser, H-R.~Sinning, Internal Friction in Metallic
  Materials: A Handbook,  Springer--Verlag (Berlin,  2007).
%
\end{thebibliography}
\end{document}